\documentclass[twocolumn,aps]{revtex4}
\usepackage{graphicx} 
\usepackage{hyperref}
\usepackage{natbib}
\usepackage[papersize={8.5in,11in},margin=1
in]{geometry}
\usepackage{tabularx}
\usepackage{float}
\usepackage{multirow}
\usepackage{array}
\usepackage{subcaption}

\hypersetup{
    colorlinks, 
    linkcolor=black,
    urlcolor=blue,
    citecolor=black}

\begin{document}
\title{MKS 390 Micro-Ion Gauge Performance After Exposure to High Pressures}
\author{B. Massett\textsuperscript{1}, W. T. Shmayda\textsuperscript{2}}
\affiliation{Rochester Institute of Technology\textsuperscript{1}, Tritium Solutions Inc.\textsuperscript{2} \\ January 29, 2025}

\begin{@twocolumnfalse}
    \centering
    \begin{abstract}
    \vspace{15mm}
    \begin{center}
    \noindent\rule{14.5cm}{0.4pt}
    \end{center}
    \section*{Abstract}
In certain applications, pressure transducers may be exposed to high pressures either deliberately or accidentally, raising concerns about their functionality afterwards. We compared the performance of two MKS Granville-Phillips 390 Micro-Ion Gauges against each other, one that had been exposed to 10,000 Torr and the other had never been exposed to pressures above 1000 Torr. Our results show that the differences in the readings between the gauges were within the range of uncertainty specified by the manufacturer indicating negligible impact due to the exposure to high pressure.  Additionally, the high pressure exposure did not compromise the leak integrity of the gauge.
      
    \vspace{3mm}
    \begin{center}
    \noindent\rule{14.5cm}{0.4pt}
    \end{center}
    \vspace{5mm}
      
    \end{abstract}
\end{@twocolumnfalse}

\maketitle

\vspace{20mm}
\section*{Introduction}
    The MKS Granville-Phillips 390 Micro-Ion Gauge is a wide-range pressure transducer for use in high-vacuum systems. The gauge contains four separate pressure sensors, which are sensitive to different pressure ranges: two piezo-resistive diaphragm sensors, one which acts as a reference and the other which measures the vacuum pressure, a mid-range thermal conductivity pressure sensor which relies on heat loss from a gold-plated tungsten wire, and a high-vacuum Bayard-Alpert ionization gauge (BAG) which generates a current proportional to the gas density.  The BAG operates at two emission currents: 20 µA when the pressure exceeds $5\cdot 10^{-6 }$ Torr and 4 mA below that value. The thermal conductivity sensor turns on when the pressure decreases to 20 mTorr. The piezo-resistive diaphragm sensor operates from atmosphere down to 20 mTorr. \citep{MKS}. 

    MKS specifications indicate that the overall accuracy including the performance of the gauge, the sensors, and the electronics for air is as follows: 15\% of the reading for hard vacuum up to 100 mTorr, 10\% from 100 mTorr to 150 Torr and 2.5\% from 150 Torr to 1000 Torr.  The corresponding repeatability for those ranges is 5\%, 2.2\% and 1\% of the reading respectively. An absolute maximum pressure is not provided by MKS although, a maximum value of 1000 Torr is implied by the specification. Excluding a containment rupture, inspection of the gauge cross section suggests that only the piezo-resistive diaphragm sensors would be sensitive to damage at elevated pressures. When Pisana at Torion Plasma exposed one module to 10,000 Torr, the gauge exhibited no apparent damage. \citep{Pisana}.

    The authors of this paper have examined the Pisana gauge exposed with the objective of measuring the performance of the 'exposed' gauge against a similar off-the-shelf unit, the 'control' gauge, that had never been exposed to pressures in excess of one atmosphere.

    \vspace{-3mm}
    
\section*{Experimental Setup}
    We built a simple vacuum system to test the performance of both the control and the exposed pressure gauges. 

    \begin{figure}[ht]
        \centering \includegraphics[width=0.7
        \columnwidth]{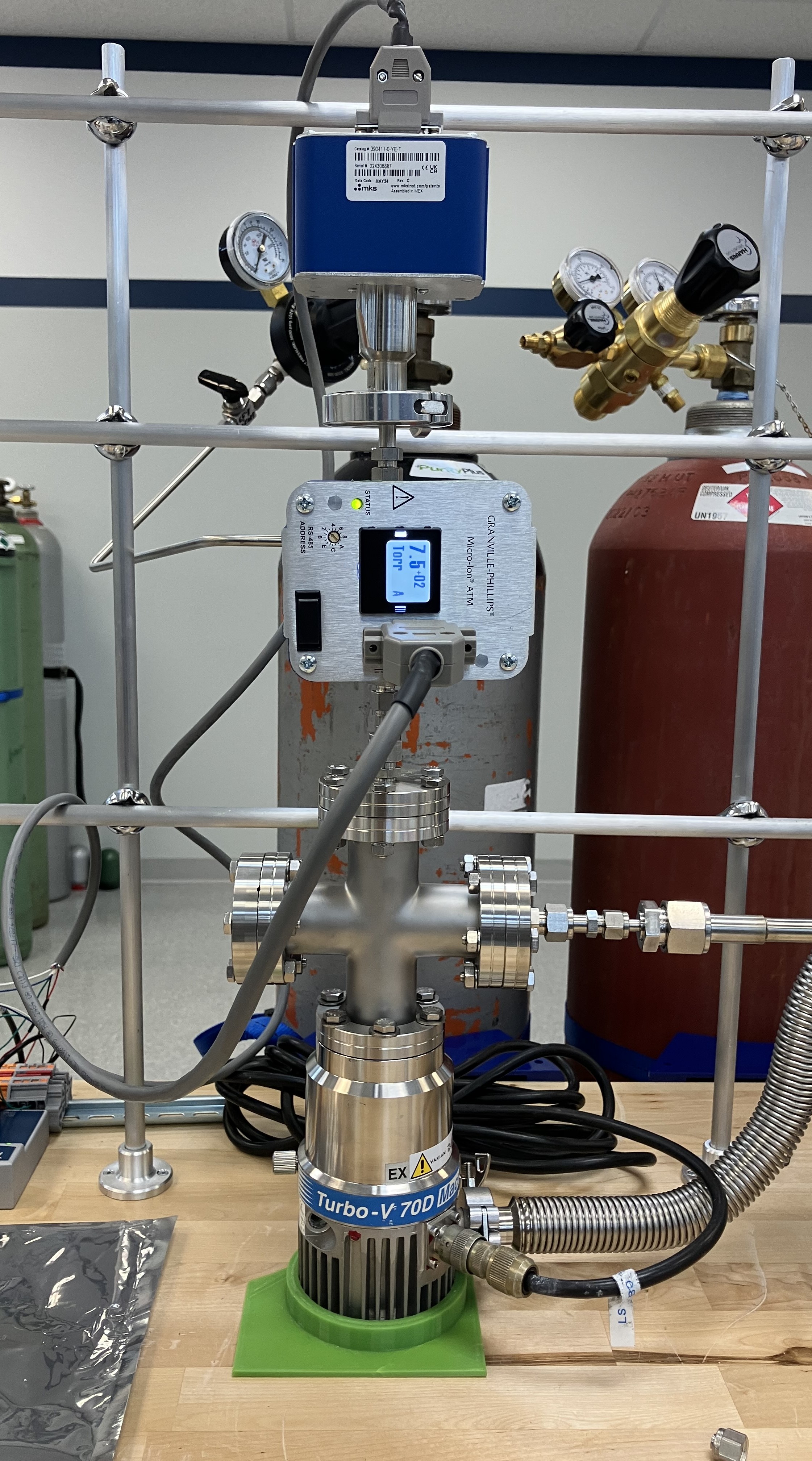}
        \caption{Photograph of the Vacuum system setup used to compare the two MKS 390 pressure transducers}
        \label{fig:Picture}
    \end{figure}

\begin{figure}[ht]
        \centering \includegraphics[width=1.2\columnwidth]{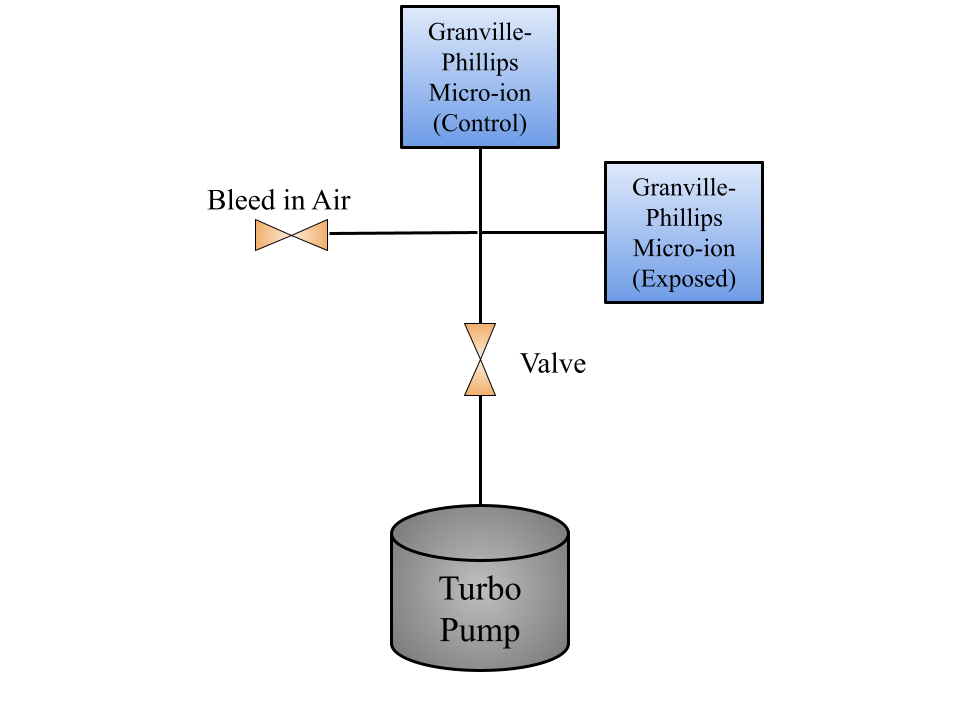}
        \caption{Diagram of the vacuum system used to assess the performance of the  pressure transducers}
        \label{fig:Diagram}
    \end{figure}

    A photo of the vacuum system is provided in Fig. \ref{fig:Picture}. The stainless steel system comprised an IDP-3 scroll roughing pump connected to the Varian Turbo-V 70D vacuum pump. A valve was installed between the pump set and the transducer assembly to isolate the transducers from the pump train so that the upper portion of the loop could be pressurized. The setup was helium leak-tight to $1.5 \cdot 10^{-9 } atm\cdot cc/s$. A cartoon of this vacuum system is provided in Fig. \ref{fig:Diagram}.
    
    The transducers were tested by first evacuating the whole system with the turbo pump to $1\cdot 10^{-6 }$ Torr. Once the pressure stabilized, we isolated the transducers from the pump train and monitored the subsequent pressure increase on both transducers through the low pressure regime. 

    To access pressure above a few torr, a valve in the assembly was cracked open to slowly bleed air into the system to increase the pressure in a controlled manner in steps of approximately 50 Torr up to one atmosphere. At each step, the pressure was held for a few minutes to get a stable response from both devices. The analog voltage output from both gauges was averaged over the time that the pressure was held constant.

    After completion of the initial tests, the control gauge was temporarily removed from the setup. The exposed gauge underwent ten more high-pressure cycles up to a maximum of 10 atm. After this, more testing was performed on both gauges to ensure the repeatable resilience of the transducers to high pressure.

\section*{Results}

\begin{figure}[ht]
        \centering \includegraphics[width=1.0\columnwidth]{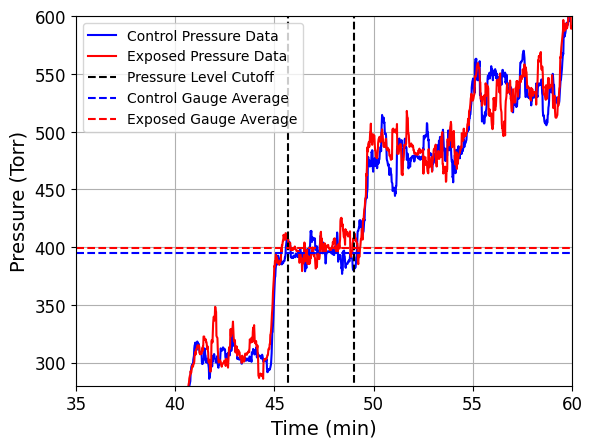}
        \caption{Typical step profiles used to scan through a range of system pressures to determine the mean pressure at each pressure setting}
        \label{fig:Example}
    \end{figure}

    Fig. \ref{fig:Example} shows a typical step profile over the pressure range 300 Torr to 550 Torr for the two transducers. The dwell time at each pressure level was arbitrary but long enough to provide good statistics.  The pressure reading for each transducer was averaged over the dwell time. The complete pressure scan is provided in the Appendix, refer to Fig. \ref{fig:Full}.  

\begin{figure}[ht]
        \centering \includegraphics[width=1.0\columnwidth]{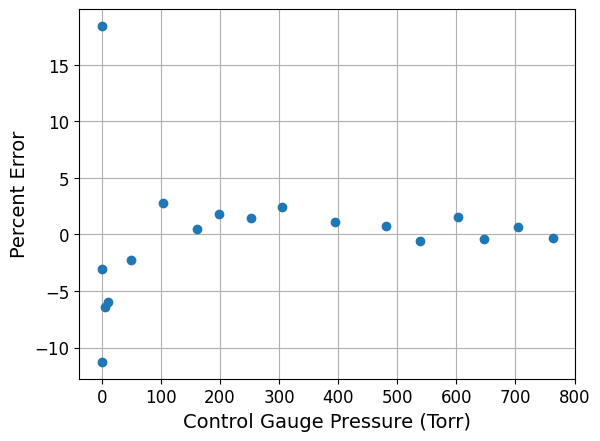}
        \caption{Percentage discrepancy between the exposed transducer and the control gauge as a function of pressures ranging from 50 to 760 Torr}
        \label{fig:Error}
    \end{figure}

    Inspecting Fig. \ref{fig:Example} shows that the readings provided by the two gauges appear to coincide.  To quantify the difference in readings between the two gauges, the average reading of the exposed gauge was subtracted from the average reading of the control gauge at each pressure setting and that difference was divided by the average reading of the control gauge. This percentage difference was plotted against the pressure values from the control gauge for the pressure range spanning from $2.26\cdot10^{-4}$ Torr to $760$ Torr. The results are provided in Fig. \ref{fig:Error} and indicate that the discrepancy between the two gauges is on the order of 2\% from 100 Torr to 760 torr and appears to be randomly distributed about 0\% above 400 Torr. At pressures below 400 Torr and dropping to '0' Torr, the discrepancy increases to about 10\%.

\begin{figure}[ht]
        \centering \includegraphics[width=1.0\columnwidth]{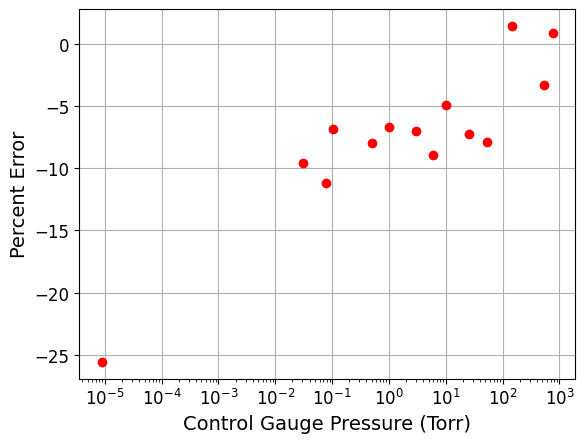}
        \caption{Percentage discrepancy between the exposed transducer and the control gauge as a function of pressures ranging from 0.02 to 80 Torr}
        \label{fig:Error2}
    \end{figure}

    A second data set was collected over a broader pressure range spanning from $8.9\cdot10^{-6}$ Torr to $760$ Torr to investigate the discrepancy at pressures below 100 Torr in greater detail. The analysis of this data set is provided in Fig. \ref{fig:Error2}.  This data indicates that the discrepancy appears to be randomly distributed around -7.5\% suggesting that the exposed gauge reads lower on average than the control gauge in the pressure range $2\cdot10^{-2}$ Torr to 80 Torr. The discrepancy increases to -25\% at $1\cdot10^{-5}$ Torr at the lowest pressure. The complete pressure scan is provided in Fig. \ref{fig:Full2} in the Appendix.

    We then took the exposed gauge to high pressures again after the analysis of these two data sets showed promising results for its performance. The exposed gauge underwent ten additional pressure cycles with the peak pressure being approximately 10 atm. We did two more performance tests after these pressure cycles to ensure the repeatable exposure to high pressure did not affect the transducer.

\begin{figure}[ht]
        \centering \includegraphics[width=1.0\columnwidth]{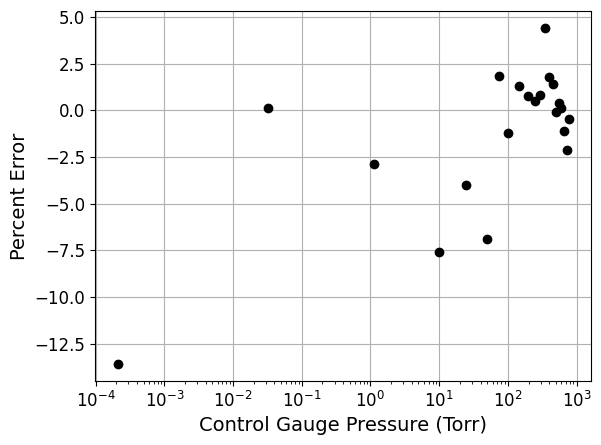}
        \caption{Percentage discrepancy between the exposed transducer and the control gauge as a function of pressure after 10 additional pressure cycles}
        \label{fig:Error3}
    \end{figure}

    The first data set collected after high pressure cycles spanned from $1.6\cdot10^{-4}$ Torr to $760$ Torr. The analysis of this data set is provided in Fig. \ref{fig:Error3}.  This data indicates that the discrepancy appears to be randomly distributed around -5\%, again suggesting that the exposed gauge reads lower on average than the control gauge in the pressure range $2\cdot10^{-2}$ Torr to 80 Torr. The discrepancy increases to -13\% at $1\cdot10^{-4}$ Torr at the lowest pressure. The complete pressure scan is provided in Fig. \ref{fig:Full3} in the Appendix.

\begin{figure}[ht]
        \centering \includegraphics[width=1.0\columnwidth]{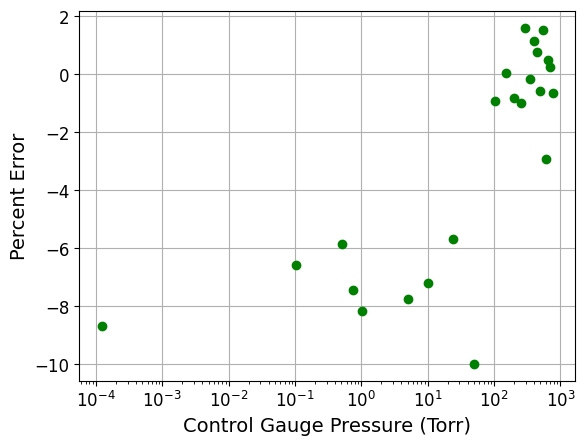}
        \caption{Percentage discrepancy between the exposed transducer and the control gauge as a function of pressure}
        \label{fig:Error4}
    \end{figure}

    The second data set collected after high pressure cycles spanned from $9.8\cdot10^{-5}$ Torr to $760$ Torr. The analysis of this data set is provided in Fig. \ref{fig:Error4}.  This data continues to show a discrepancy where the exposed gauge reads lower on average than the control gauge in the pressure range $2\cdot10^{-2}$ Torr to 80 Torr. The discrepancy here is again randomly distributed around -7.5\%, included on the lowest pressure data point. The complete pressure scan is provided in Fig. \ref{fig:Full3} in the Appendix.

\section*{Uncertainty Propagation}

    To understand the percent discrepancy between the exposed transducer and the control, we applied uncertainty propagation theory to the two data sets: 

    \begin{equation}
        \delta f = \sqrt{\sum({\frac{\partial f}{\partial x}})^2(\delta x)^2}
        \label{eq:Error}
    \end{equation}

    Expanding Eq. \ref{eq:Error}, the uncertainty propagation in the pressure difference between the two transducers becomes:

    \begin{equation}
        \delta P_{Diff} = \sqrt{(\delta P_{control})^2 + (\delta P_{exposed})^2}
        \label{eq:ErrorPdiff}
    \end{equation}

    where $\delta P_{control}$ is the uncertainty in the measurement using the control transducer and $\delta P_{exposed}$ is the uncertainty in the measurement using the exposed transducer. The measurement uncertainties for each pressure range are listed in the manual \citep{MKS} and have been summarized in the introductory section of this document.

    \begin{equation}
        \delta (\%Err) = 100\cdot\sqrt{(\frac{\delta P_{Diff}}{P_{control}})^2 + (\frac{P_{Diff}\delta P_{control}}{P_{control}^2})^2}
        \label{eq:PercentError}
    \end{equation}
    
 \begin{figure}[ht]
        \centering \includegraphics[width=1.0\columnwidth]{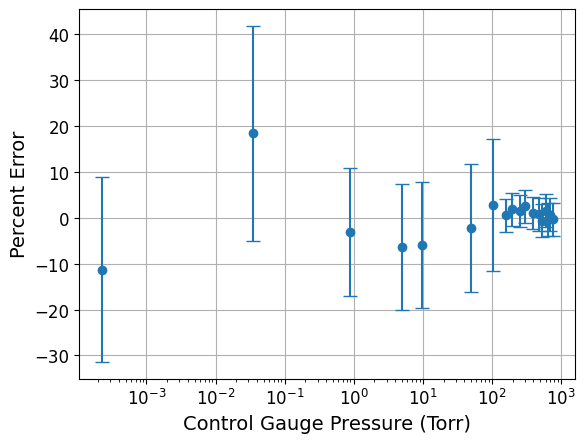}
        \caption{Expected percentage uncertainty in the pressure measurements between two MKS 390 as a function of pressure from $1\cdot10^{-4}$ to $760$ Torr}
        \label{fig:UncertaintyError}
    \end{figure}
    
\begin{figure}[ht]
        \centering \includegraphics[width=1.0\columnwidth]{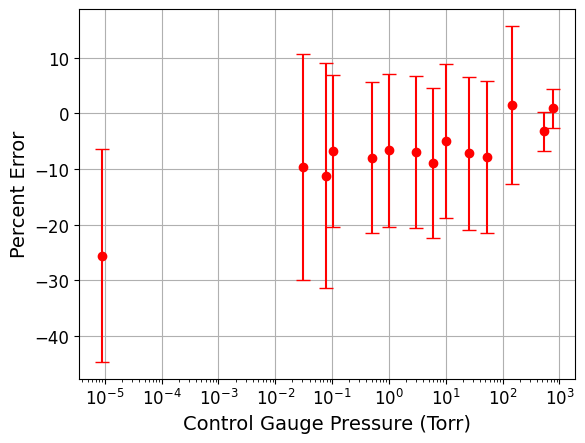}
        \caption{Expected percentage uncertainty in the pressure measurements between two MKS 390 as a function of pressure from $1\cdot10^{-5}$ to $760$ Torr}
        \label{fig:UncertaintyError2}
    \end{figure}

    Equation \ref{eq:PercentError} defines how the uncertainties in the readings of the two gauges should combine to form an uncertainty in the overall estimate of the discrepancy. These values are plotted in Fig. \ref{fig:UncertaintyError}. Inspection of this figure reveals that the uncertainty is on the order of +/- 5\% in the 100 to 760 Torr range and increases to +/- 10\% below 100 Torr and increases further to +/- 20\% at pressures below 0.1 Torr as seen in Fig.\ref{fig:UncertaintyError2}. Inspection of figures Fig.\ref{fig:Error} and Fig. \ref{fig:Error2} indicates that the discrepancies between the two gauges lie well within the uncertainties provided by the manufacturer. The measurement differences observed between the two gauges over the examined pressure range are not statistically significant.

\begin{figure}[ht]
        \centering \includegraphics[width=1.0\columnwidth]{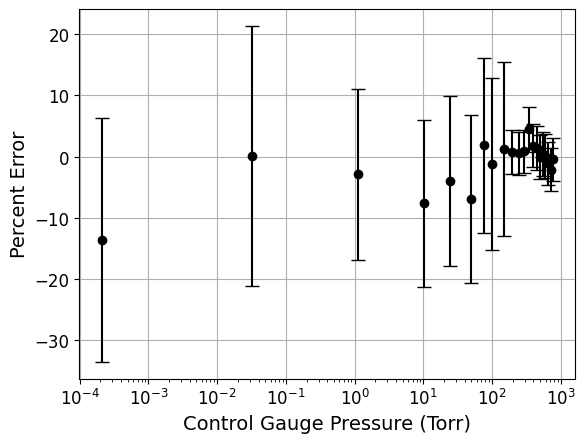}
        \caption{Expected percentage uncertainty in the pressure measurements between two MKS 390 as a function of pressure from $1\cdot10^{-4}$ to $760$ Torr after 10 additional pressure cycle}
        \label{fig:UncertaintyError3}
    \end{figure}

\begin{figure}[ht]
        \centering \includegraphics[width=1.0\columnwidth]{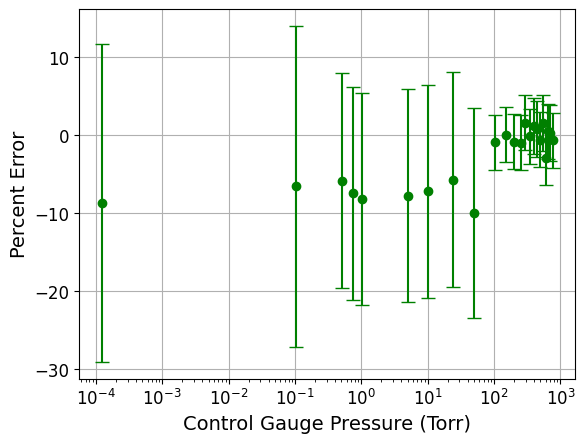}
        \caption{Expected percentage uncertainty in the pressure measurements between two MKS 390 as a function of pressure from $1\cdot10^{-4}$ to $760$ Torr after 10 additional pressure cycle}
        \label{fig:UncertaintyError4}
    \end{figure}

    The same uncertainty analysis was performed on the data sets collected after the additional high pressure cycling. The analysis of the first trial after the cycling is seen in Fig. \ref{fig:UncertaintyError3}, while the analysis of the second trial after cycling is seen in Fig. \ref{fig:UncertaintyError4}. Both of these trials reveal the same uncertainty values in the same ranges as seen in the previous two trials. The measurement differences again fall within these uncertainty ranges and are not statistically significant.

\section*{Conclusion}
    Pressure readings from an MKS 390 Micro-Ion gauge that was over-pressurized by a factor of 10 to 10,000 Torr were consistent with a similar gauge that had not been over-pressurized within statistical error.  Over-pressurizing the MKS 390 gauge did not adversely affect its ability to measure pressures within the manufacturer's uncertainty ranges. At pressures above 100 Torr, the discrepancy between the gauges was on the order of, or less than, 2.5\%. The largest deviation occurred at lower pressures, peaking at -25\% at a pressure of $8.86\cdot10^{-6}$ Torr.
   
    With the possible exception of a loss of vacuum integrity the MKS 390 Micro-Ion gauge is capable of withstanding a pressure excursion as high as 10,000 Torr.

\vspace{-5mm}

\section*{Acknowledgements}
This experiment was funded in part by Torion Plasma Corporation.  The authors gratefully acknowledge the contribution. 

\clearpage

\onecolumngrid

\section*{Bibliography}
\vspace{-8mm}
\bibliographystyle{unsrt}
\bibliography{citations.bib}

\begin{thebibliography}{1}

\bibitem{MKS}
MKS Pressure and Vacuum Measurement Solutions.
\newblock {\em Micro-Ion®ATM Module Instruction Manual - 390001}.

\bibitem{Pisana}
Simone Pisana.
\newblock Pressure tests mks granville-phillips 390 micro-ion atm vacuum gauge.
\newblock Technical report, Torion Plasma Corporation, 2024.

\end{thebibliography}

\clearpage

\onecolumngrid

\section*{Appendix}

    \begin{figure}[ht]
    \centering
    \noindent\rule{\columnwidth}{0.4pt}
    \begin{subfigure}[b]{0.45\textwidth}
    \includegraphics[width=\textwidth]{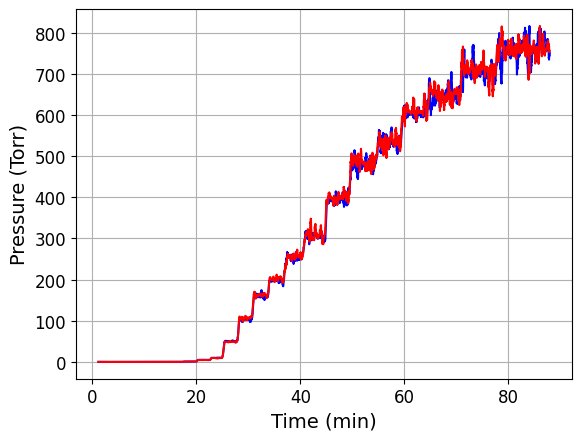}
        \caption{Full pressure scan with a focus on the high pressure regime plotted linearly}
        \label{fig:Full_Linear}
    \end{subfigure}
    \hfill
    \begin{subfigure}[b]{0.45\textwidth}
    \includegraphics[width=\textwidth]{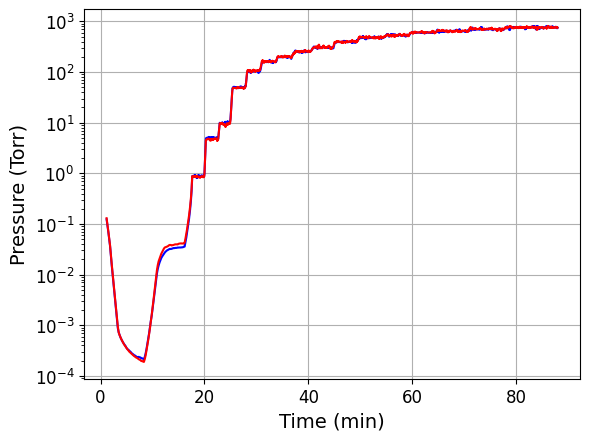}
        \caption{Full pressure scan with a focus on the high pressure regime plotted logarithmically}
        \label{fig:Full_Log}
    \end{subfigure}
    \noindent\rule{\columnwidth}{0.4pt}
    \caption{Full pressure data set}
    \label{fig:Full}
    \end{figure}

    \begin{figure}[ht]
    \centering
    \noindent\rule{\columnwidth}{0.4pt}
    \begin{subfigure}[b]{0.45\textwidth}
    \includegraphics[width=\textwidth]{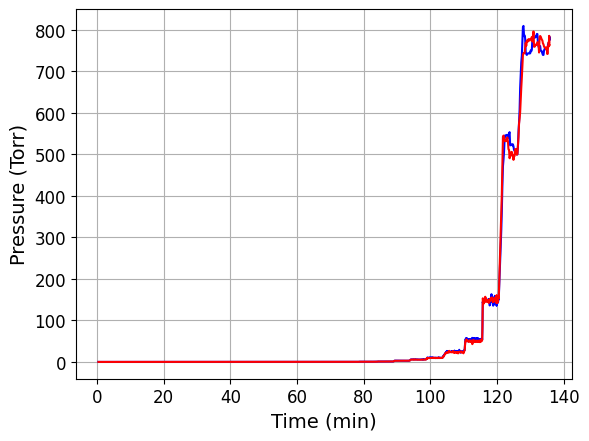}
        \caption{Full pressure scan with a focus on the low pressure regime plotted linearly}
        \label{fig:Full_Linear2}
    \end{subfigure}
    \hfill
    \begin{subfigure}[b]{0.45\textwidth}
    \includegraphics[width=\textwidth]{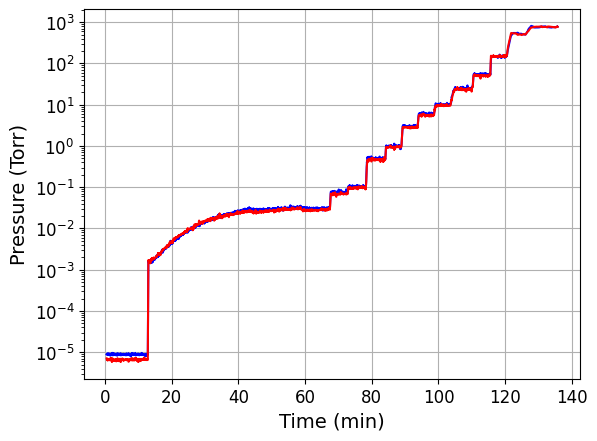}
        \caption{Full pressure scan with a focus on the low pressure regime plotted logarithmically}
        \label{fig:Full_Log2}
    \end{subfigure}
    \noindent\rule{\columnwidth}{0.4pt}
    \caption{Full pressure data set}
    \label{fig:Full2}
    \end{figure}
  \newpage

  \begin{figure}[ht]
    \centering
    \noindent\rule{\columnwidth}{0.4pt}
    \begin{subfigure}[b]{0.45\textwidth}
    \includegraphics[width=\textwidth]{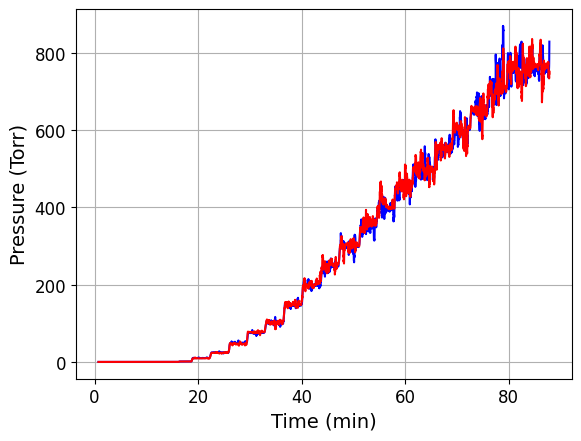}
        \caption{Full pressure scan plotted linearly}
        \label{fig:Full_Linear3}
    \end{subfigure}
    \hfill
    \begin{subfigure}[b]{0.45\textwidth}
    \includegraphics[width=\textwidth]{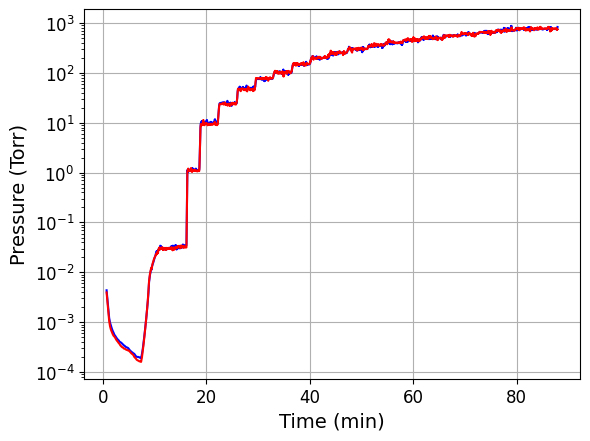}
        \caption{Full pressure scan plotted logarithmically}
        \label{fig:Full_Log3}
    \end{subfigure}
    \noindent\rule{\columnwidth}{0.4pt}
    \caption{First full pressure data set after additional high-pressure cycles}
    \label{fig:Full3}
    \end{figure}

    \begin{figure}[ht]
    \centering
    \noindent\rule{\columnwidth}{0.4pt}
    \begin{subfigure}[b]{0.45\textwidth}
    \includegraphics[width=\textwidth]{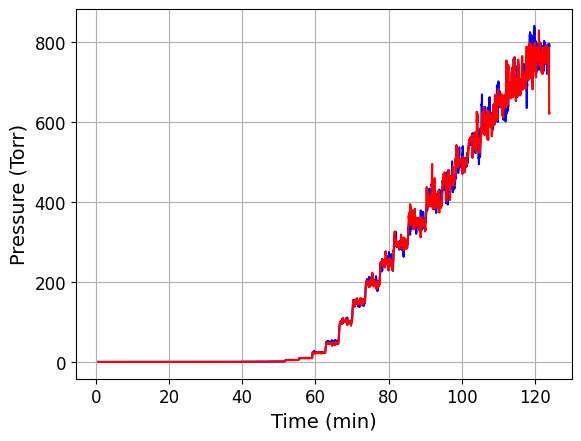}
        \caption{Full pressure scan plotted linearly}
        \label{fig:Full_Linear4}
    \end{subfigure}
    \hfill
    \begin{subfigure}[b]{0.45\textwidth}
    \includegraphics[width=\textwidth]{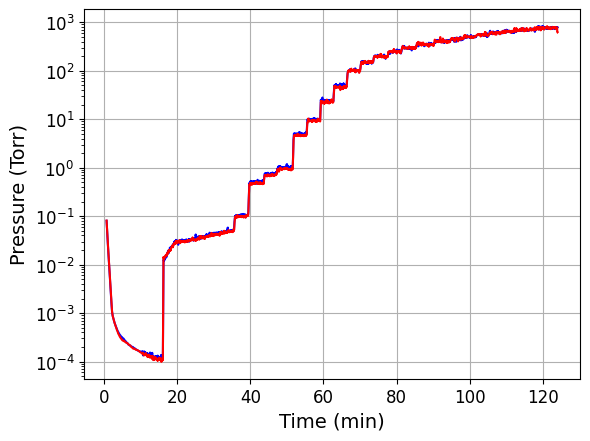}
        \caption{Full pressure scan plotted logarithmically}
        \label{fig:Full_Log4}
    \end{subfigure}
    \noindent\rule{\columnwidth}{0.4pt}
    \caption{Second full pressure data set after additional high-pressure cycles}
    \label{fig:Full4}
    \end{figure}
    \newpage

\end{document}